# What is the Point of Fairness? Disability, AI and The Complexity of Justice

**Cynthia L. Bennett**
Human Centered Design & Engineering
University of Washington
Seattle, WA, United States
bennec3@uw.edu

**Os Keyes**
Human Centered Design & Engineering
University of Washington
Seattle, WA, United States
okeyes@uw.edu

**ABSTRACT**
Work integrating conversations around AI and Disability is vital and valued, particularly when done through a lens of fairness. Yet at the same time, analyzing the ethical implications of AI for disabled people solely through the lens of a singular idea of "fairness" risks reinforcing existing power dynamics, either through reinforcing the position of existing medical gatekeepers, or promoting tools and techniques that benefit otherwise-privileged disabled people while harming those who are rendered outliers in multiple ways. In this paper we present two case studies from within computer vision - a subdiscipline of AI focused on training algorithms that can "see" - of technologies putatively intended to help disabled people but, through failures to consider structural injustices in their design, are likely to result in harms not addressed by a "fairness" framing of ethics. Drawing on disability studies and critical data science, we call on researchers into AI ethics and disability to move beyond simplistic notions of fairness, and towards notions of justice.

**INTRODUCTION**
As machine learning becomes more ubiquitous, questions of AI and information ethics loom large in the public imagination. Much concern has been focused on promoting AI that results in more fair outcomes that do not discriminate against protected classes, such as those marginalized on the basis of gender and race. Yet little of that work has specifically investigated disability. Two notable exceptions, both from within the spaces of Disability Studies and Assistive Technology (AT), are Shari Trewin's statement on "AI Fairness for People with Disabilities" [27], and the World Institute on Disability's comments on AI and accessibility [32]. Together they argue that making disability explicit in discussions of AI and fairness is urgent as the quick, black boxed nature of automatic decision-making exacerbates disadvantages people with disabilities already endure and creates new ones. Though low representation in datasets is blamed, increasing representation will be complex given disability politics. For example, disabled people strategically choose whether and how to disclose their disabilities (if they even identify as having disabilities), likely leading to inconsistent datasets even when disability information is intentionally collected. Additionally, disabilities present themselves (or not) in a myriad ways, destabilizing (category-dependent) machine learning as an effective way of correctly identifying them.

We are encouraged by the nascent engagement between Disability Studies, AT and AI ethics, and agree with many of the concerns outlined in both documents. For example, healthcare and employment remain out of reach for many disabled people despite policies that prohibit discrimination on the basis of disability, and we would be remis to deny AT's role in increasing quality of life for some people with disabilities, if incremental. At the same time, "fairness", like "equality" [1], is not an uncontested concept. Ethicists have troubled the notion that it can produce justice in and of itself. A recent paper by Anna Lauren Hoffmann, for example [12], pointing to the way that fairness is modelled on U.S. anti-discrimination law, surfaces the gaps and injustices a fairness framing remains silent on, including its failure to dismantle and rework structural oppression. In fact, without addressing the hierarchies that disadvantage people with disabilities in the first place, Hoffmann and disability justice activists argue, fairness may reproduce the discrimination it seeks to remedy. Justice, on the other hand, guides recovery aimed at repairing past harm and may scaffold more accountable and responsible AI that is equitable in its handling of data as well as deployment (or withholding). Therefore, we argue for a reframing from fairness to justice when concerning AI ethics and disability.

To highlight the necessity of this reframing, we present two case studies of AI/AT in which the application of a principle of fairness, while an improvement over inaction, does not prevent the harms that the technology opens space for.

**FAIRNESS: DEFINED AND CONTESTED**
To begin, we briefly define and critique fairness according to the scoped scholarship on AI, ethics and people with disabilities that we use to form our arguments. Shari Trewin and Anna Laura Hoffmann summarise fairness as it has been articulated by statisticians. These articulations (as applied to disability) largely evaluate whether similar cases, separated only by the presence of disability, produce the same outcome. In the event there is a disparity, the system is considered unfair. Approaches to remedying this assume that a lack of fairness is a failure of *implementation*, and largely center increasing the representativeness of the data underlying the algorithm, and improving how the algorithm integrates "outliers"[29, 12]. An initial reading of fairness may make it seem a reasonable goal to center in algorithmic systems; after all, what is the problem with addressing disparities?

Hoffmann delineates four primary limitations of an approach to the consequences of AI that centers fairness. First, fairness is not fair; it aims to increase the status of disadvantaged people without explicitly addressing how the privileges of more powerful people will change. Second, fairness relies on traits being well defined so a system can know what to do with them. Third, fairness historically aims to improve one contested identity at a time when in reality, many people are multiply marginalized and it is not just aspects of their identity but interactions among multiple facets of their identities and oppressive structures which produce systematic discrimination in different ways in different situations. Finally, fairness frames marginalization as occurring in relation to specific things, namely assumed desires like employment. However, much injustice is produced by the development of standards (both formally and informally) which are then applied across domains. In other words, someone is not marginalized when applying for jobs and then not. Instead, oppression is threaded through what they do according to predetermined norms and disciplinary institutions which enforce them. In summary, fairness is premised on understanding how oppression manifests in an individual and aims to promote equality through the remediation of technologies. It does not question the structures (including those which rely on AI to surveil and make decisions) that allow people with disabilities to be disadvantaged in the first place.

For the rest of this paper, we will present two case studies—one on the use of AI to diagnose neurodiversity including autism and the second about computer vision which provides information for blind people. after introducing a case, we will overview some concerns that might be raised through a fairness lens and then some concerns which might be raised with a justice lens. Through these cases, we hope to concretize differences between the two lenses and demonstrate how justice can situate and pluralize our conversations on AI, ethics, and disability to address societal, structural oppression beyond improving automatic decision-making and datasets themselves.

## CASE STUDIES OF AI IN ASSISTIVE TECHNOLOGY

### AI For Diagnosis

A body of research within computer vision attempts to create systems which can (using facial recognition) automatically diagnose certain neurodiverse states – including autism [27, 23]. Using already-recognized and diagnosed autistic children, researchers rely on examining facial expressions, degrees of emotiveness, and repetitive behaviours to provide diagnostic tools, arguing that doing so would reduce the delay of diagnosis in a child's life [22, 11].

### CONCERNS RAISED THROUGH A FAIRNESS LENS

With diagnosis, researchers are confronted with biases in the pre-existing framework of autism – particularly the widely-studied gender bias in symptoms [24], and the consequential discrepancies in diagnostic rates [10] – and less-studied but firmly established biases around race and ethnicity, class and geography [4, 16]. Dependence on diagnostic tools which are based on the experiences of *those already diagnosed* thus risk replicating these biases, providing seemingly-objective rigor to determinations that a child presenting inconsistently with (white, assigned male at birth) autistic children cannot be autistic, and should be gatekept out of support systems. With a fairness metric, we might suggest diversifying datasets so marginalized genders and races can be correctly diagnosed. But this solution may not adequately consider what it means to have the power to diagnose, and who might endure what consequences as a result.

### CONCERNS RAISED THROUGH A JUSTICE LENS

In the case of diagnostic tools for autism, we run into concerns around medicalization and gatekeeping: the distinct power that comes with diagnostic authority given the institutionalization of a medical model of disability into the power structures of society [7].

Tools to "help" autistic people in the model of existing computer vision prototypes do not just provide diagnosis – they also reinforce the notion that the formal diagnostic route is the only legitimate one for autistic existences, reinforcing the power that psychiatrists hold. Examinations of medicalization – the process by which this notion of formal gatekeeping becomes legitimized – have already identified it within autism diagnostics [31], simultaneously finding little validity to the diagnostic systems computer vision researchers are using as their baseline [28]. By adding technical and scientific authority to medical authority, people subject to medical contexts are not only not granted power, but are even further disempowered, with even less legitimacy given to the patient's voice. Once again, fairness is not a solution; the issue is not one of discrimination against the patient for being autistic but for being a *patient*. Just outcomes in this area, in other words, require not a consideration of fairness but of power, and of the wider social context into which technical systems are placed.

Finally, and more cut-and-dried, there is the question of what the consequences and implications of an autism diagnosis are. AI systems in this domain are built on the premise that an early diagnosis is a good outcome, that diagnosis leads to possibilities of treatment, support and consideration. Notwithstanding the already-discussed biases in who can access diagnosis (and how diagnostic tests are constructed), there are serious questions about whether an earlier diagnosis is a better one [25]. Rather than helping people, earlier diagnoses may harm them.

Even worse consequences stem from the fact that autism is not "just" a diagnostic label, whatever computer vision researchers may think; it is a label that carries with it certain associations about financial cost, incapability and risk – associations that have led to myriad harmful behavior change therapies and autistic children being murdered as "mercy killings"[35, 19]. As Mitzi Waltz puts it, "autism = death". Morally and ethically, computer vision systems to provide that label, if designed without attendance to the wider societal contexts in which autistic people live, might well be considered death too.

Autism diagnosed with AI *is* an issue of fairness – an issue of the unfair treatment of autistic people – but it is not one that can be solved simply through examining the immediate algorithmic inputs and outputs of the computer vision system. Instead, we need a model that considers holistic, societal implications, and the way that technologies alter the life chances of those they are used by or on.

### AI For "Sight"

Our second case concerns a longstanding area of research – engaged in by AI researchers, health researchers, and HCI researchers – is that of using computer vision (AI that "sees") to assist vision-impaired people [33, 14, 19]. Projects presented at ASSETS alone include haptic/vision-based systems for detecting and representing the emotions behind facial expressions [5], augmented touch for communicating visual information [9], facial recognition for communicating conversational partners' identity [2], and object and scene recognition [20].

### CONCERNS RAISED THROUGH A FAIRNESS LENS

First, we must ask: sight for whom and what gets seen? There is a longstanding recognition of biases within computer vision systems, and limitations in their ability to represent the complexity of the world – biases that often impact those already marginalized [13]. In the case of object recognition, for example, a recent paper demonstrates that such systems are developed largely in a white, western and middle-class context, failing to recognize common household objects that are more-often found in poor or non-western environments [8]. The centering of such systems in AT design risks further harm to people already marginalized within both society widely, and the disability community. And improving algorithms to recognize more genders, races and objects still predispose futures where surveillance technologies may be justified for their utility for blind people while ignoring their ongoing documented misuse [12, 13].

### CONCERNS RAISED THROUGH A JUSTICE LENS

Unlike AI for diagnosis, computer vision to help people see seems to put more control in the disabled users hands. They are not the focus of the gaze: they are gazing. But this inversion does not necessarily redistribute power in a positive fashion; it can still promote asymmetric and harmful power distributions. Whereas tools like a white cane assume the brain the analytical unit, computer vision may transfer such judgment to automatic decision-making. Though developers of many identifying technologies clarify that their use is meant to support not replace human decision-making, we know that technology is often pedestalized; that technological and scientific ways of "knowing" are treated as superior to the alternative, and frequently deferred to even in the presence of contradictory information or assumed to be far more accurate than they are [15]. The result is that a computer vision system for accessibility, while rendering things more accessible, does so by shifting the center of analysis and judgment away from the user and towards the (frequently expensive, black-boxed and commercially shaped) technology in hand.

Finally, computer vision, even deployed fairly, cements vision as a superior sense and legitimizes surveillance. Much research, including that cited to inform AI for accessibility, acknowledge and even praise nonvisual sensemaking. Accessibility researchers hardly advocate substituting this knowledge with technology. Yet these gestures would be more substantive if the same rigor and enthusiasm were applied to the development of technologies which train in or privilege nonvisual sensemaking [34]. Next, surveillance technologies are controversial, and disability studies scholars have critiqued the ubiquity and inaccuracy of technology savior narratives which hail automation for increasing the quality of life of people with disabilities [30]. Here we risk glorifying surveillance without questioning its misuse. How could technology to assist a blind person be kept from integration into policing technologies; who's to say blind people aren't among the users of policing technologies? Instead, until significant work is done to correct and nuance stories about disability, those who question using surveillance technologies even when used for the purposes of assisting disabled people may be shamed [21].

These are not issues that notions of "fairness" can surface, articulate and tackle, because the issue is not only that disabled sub-populations may be treated unequally between each other or compared to normative society, but that the technologies' model of liberation is liberation without challenging wider structures of power.

### CONCLUSION

We have presented two case studies of AI interventions in disabled lives, and the issues they raise around and with fairness. As we have made clear, we believe that fairness – a concept already being shifted away from in critical data studies – is highly dangerous for conversations around disability and AI to center. Rather, we advocate that everyone interested in questions of disability and AI critically examine the overarching social structures we are participating in, upholding and creating anew with our work. Doing so requires and results in a centering of our work not on questions of fairness, but on questions of *justice* [12].

There are many places to draw from in doing that. technology has always been a part of the construction of disability, and of the nature of disabled lives. Consequently, Disability Studies has long-considered questions of technology, and continues to do so. Just as Mankoff *et al.* urged the integration of disability studies into assistive technology [17], we urge a similar integration of AI and Disability conversations with Disability Studies conversations around technology, justice and power – conversations that are already taking place [3].

Similarly, though disability alone leads to unique life experiences and oppression, there is myriad scholarship on AI and black lives, trans lives, poor lives – and many of those lives are disabled lives, too. As such, it is imperative that efforts concerning just developments and deployments of AI for people with disabilities center multiply marginalized disabled people, or we risk only helping the most privileged. Additionally, we need to carve out space in AI ethics programs which are not considering disability, calling in the disability forgetting that has gone on in many purportedly justice-oriented conversations. AI is new – but the systems of oppression that give rise to disability are very, very old. They will not be unraveled piecemeal, or separate from recognizing and reckoning with the structural inequalities that have made unjust AI possible.

### ACKNOWLEDGMENTS
Our thanks to Margret Wander, Nikki Stevens, Anna Lauren Hoffmann and Adam Hyland for their feedback, help and support.